# Engineering topological edge states in two dimensional magnetic photonic crystal

Bing Yang,[1, 2] Tong Wu,[1] and Xiangdong Zhang[1,a]

[1]*Beijing Key Laboratory of Nanophotonics & Ultrafine Optoelectronic Systems, School of Physics, Beijing Institute of Technology, 100081, Beijing, China*

[2]*School of Physical Science and Information Engineering, Liaocheng University, 252059, Liaocheng, China; Shandong Provincial Key Laboratory of Optical Communication Science and Technology, 252059, Liaocheng, China*

[a] Corresponding author: zhangxd@bit.edu.cn

Based on a perturbative approach, we propose a simple and efficient method to engineer topological edge states in two dimensional magnetic photonic crystals. The topological edge states in the microstructures can be constructed and varied by altering the parameters of the microstructure according to the field-energy distributions of the Bloch states at the related Bloch wave vectors. The validity of the proposed method has been demonstrated by exact numerical calculations through three concrete examples. Our method makes the topological edge states "designable".

Analogous to electron topological insulators,[1,2] the study of topological edge states (TESs) in photonic microstructures has received increasing amount of attention in recent years.[3] Due to the lack of backscattering, such edge states are expected to have potential applications for future optical devices.[4] Up to now, optical TESs have been shown to exist in many photonic configurations, such as in magnetic photonic crystals (MPCs),[5-9] coupling resonators,[10-15] Floquet photonic lattices[16-17] and so on.[18-20] For a wide range of applications, people always want to obtain TESs as needed. This leads us to the question of if we can have a generic method that allows us to engineer TESs by altering the parameters of the original microstructure.

In this work, we propose a simple and efficient method to engineer TESs in two dimensional (2D) MPCs by altering parameters of the system. In this method, we combine perturbation analysis and the numerical calculations (finite element method performed with COMOSOL 5.1). The perturbation analysis provides us the guide in choosing the parameters to modify the structure. The band diagram of the structure will be calculated by using exact numerical methods. Our method is very general and can be applied to any microstructure. Thus, it opens up a way to engineer TESs.

In our analysis, we consider a 2D lattice of magnetic rods with radius $R$ immersed in air with lattice constant $a$. Under an external dc magnetic field along axis of rods ($z$ direction), the rods perform strong gyromagnetic anisotropy, with the relative permeability tensor taking the form[5,21]

$$\ddot{\mu}_r = \begin{bmatrix} \mu_r & i\kappa_r & 0 \\ -i\kappa_r & \mu_r & 0 \\ 0 & 0 & 1 \end{bmatrix}, \quad (1)$$

where $\mu_r$ and $\kappa_r$ are determined by rod component, mode frequency and external magnetic field. Here, for simplicity, we neglect effects of material dispersion and loss, assuming a constant permeability tensor with real-valued $\mu_r$ and $\kappa_r$ for a particular external magnetic field.[5,8] The relative permittivity of rod is denoted by $\varepsilon_r$, and permeability and permittivity of air background are $\mu_0$ and $\varepsilon_0$ as in vacuum, respectively. In such a MPC, the Bloch state for the electric field $\mathbf{E}_{nk}(\mathbf{r})$ at the $n$th band and wave vector $\mathbf{k}$ satisfies the Maxwell equation

$$\nabla \times \left[ \ddot{\mu}^{-1}(\mathbf{r}) \cdot \nabla \times \mathbf{E}_{nk}(\mathbf{r}) \right] = \omega_{nk}^2 \mu_0 \varepsilon_0 \varepsilon(\mathbf{r}) \mathbf{E}_{nk}(\mathbf{r}), \quad (2)$$

where the eigenfrequencies $\omega_{nk}$ give the band structures. Here $\ddot{\mu}(\mathbf{r})$ and $\varepsilon(\mathbf{r})$ are periodic functions of relative permeability and relative permittivity, taking values of $\ddot{\mu}_r$ and $\varepsilon_r$ within magnetic rods and 1 in air background in a unit cell, respectively.

The topological invariant, Chern number $C_n$, for the $n$th band is defined by[5,21,22]

$$C_n = \frac{1}{2\pi i} \int_{BZ} d^2k \left[ \nabla_k \times A_n(k) \right] \cdot \mathbf{e}_z, \quad (3)$$

with $e_z$ the unit vector along $z$ direction, and integral over the first Brillion zone. Here $A_n(k)$ is the Berry connection defined by $A_n(k) = \int_\Omega d^2r\, \varepsilon(r) E_{nk}^*(r) \nabla_k E_{nk}(r)$ for $s$ wave case (TM mode with electric field parallel to the rod axis) with integral over a unit cell $\Omega$.[5,21] The Bloch eigenfield $E_{nk}(r)$ is normalized such that $\langle E_{nk}(r)|E_{nk}(r)\rangle = 1$. Nonzero $C_n$ denotes the nontrivial topology of the band. According to the bulk-boundary correspondence theorem,[5,23] topological invariant, Chern number $C_{gap}$, for a bandgap is determined by the sum of $C_n$ of all bands below the gap $C_{gap} = \sum_n C_n$. Nonzero $C_{gap}$ indicates nontrivial topology of the bandgap and existence of the TES in the microstructure. The magnitude and sign of $C_{gap}$ determine the number and propagation direction of the TES, respectively. Closure and reopen of the bandgap (band inversion) will introduce topological phase transition and hence change the topological properties of the TES. Therefore by tuning the band structure of the 2D MPC, we can realize engineering the TES in the corresponding microstructures. In addition, from Eq. (2), we see that the band structures are depend on the functions $\ddot{\mu}(r)$ and $\varepsilon(r)$. So through altering these parameters, we can tune band structure of the MPC and hence engineer the TES. Let us denote $\tilde{\ddot{\mu}}(r)$ and $\tilde{\varepsilon}(r)$, respectively, as the new functions of relative permeability and relative permittivity after altering. Then the new eigenfrequency $\tilde{\omega}_{nk}$ can be estimated from the first-order perturbation theory,[24,25] which gives

$$\frac{\tilde{\omega}_{nk}^2}{\omega_{nk}^2} - 1 = -\frac{\int_\Omega dr\left[\tilde{\varepsilon}(r)-\varepsilon(r)\right]\varepsilon_0 |E_{nk}(r)|^2}{\int_\Omega dr\, \varepsilon(r)\varepsilon_0 |E_{nk}(r)|^2} - \frac{\int_\Omega dr\left[\tilde{\kappa}(r)-\kappa(r)\right]\left\{2\mu_0 \operatorname{Re}\left[iH_{nk,x}(r)^* H_{nk,y}(r)\right]\right\}}{\int_\Omega dr\, \varepsilon(r)\varepsilon_0 |E_{nk}(r)|^2}. \quad (4)$$

Here $\kappa(r)$ and $\tilde{\kappa}(r)$ are gyromagnetic strength functions before and after altering, $H_{nk,x}(r)$ and $H_{nk,y}(r)$ are magnetic field components scalar functions in 2D plane. For simplicity, we have set the diagonal components of permeability of all rods equal to $\mu_0$ in deriving Eq. (4).

Equation (4) provides a simple way to estimate the shift in eigenfrequency before and after changes of the structure. There are two terms in Eq. (4). The first term comes from the alteration of permittivity $\tilde{\varepsilon}(r) - \varepsilon(r)$ and the second term comes from the alteration of gyromagnetic strength in permeability $\tilde{\kappa}(r) - \kappa(r)$, and both two terms are depend on the field-energy distribution of the Bloch states in a unit cell. By altering the parameters distributions of the microstructure, we can tune the shift of eigenfrequencies. For example, if we alter the permittivity distribution by inserting isotropic rods that have positive $\tilde{\varepsilon}(r) - \varepsilon(r)$ in insertion area, the first term in Eq. (4) will get a negative value which leads to decreases of eigenfrequencies with magnitudes related to the field-energy distributions $\varepsilon_0 |\mathbf{E}_{nk}(r)|^2$. Thus in this way, Eq. (4) provides a useful guide to tune the band structures of the MPCs, even realize the band inversion, and thus engineer the TES in corresponding microstructures.

In the following we will demonstrate explicitly the validity of the above method through concrete examples. First, we consider a 2D square lattice of magnetic rods with $R=0.46a$, $\kappa_r=0.4$ and $\varepsilon_r=13$. The band diagram of the structure is shown in Fig. 1(a). It is seen clearly that there is no topological bandgap and TES in such a structure within the frequency range from $a/\lambda=0$ to 0.4. We focus on the first and second bands as marked by two bold black lines in Fig. 1(a). The Chern numbers for them are zero. Now, we inspect the field-energy distribution $2\mu_0 \text{Re}\left[iH_{nk,x}(r)^* H_{nk,y}(r)\right]$ in a unit cell for Bloch states X$_1$, X$_2$, M$_1$ and M$_2$ denoted in Fig. 1(a). The results are plotted in Fig. 1(d) with black circle denoting boundary of the rod. We see that for Bloch states X$_1$, X$_2$ and M$_1$, all the fields within rods are negative, while for state M$_2$, positive values near center of the rod lead to the integral of $2\mu_0 \text{Re}\left[iH_{nk,x}(r)^* H_{nk,y}(r)\right]$ within rod near zero (still negative). According to the perturbative guide of Eq. (4), if we increase $\kappa_r$ of rods, that is, $\tilde{\kappa}(r) - \kappa(r) > 0$ within rods, only the second term in Eq. (4) gets positive values within rods. Therefore the frequencies of Bloch states X$_1$, X$_2$ and M$_1$ will increase while the frequency of state M$_2$ will remain almost unaffected. With the increase of $\kappa_r$, the M$_1$ state will degenerate with M$_2$. In Fig. 1(b), the band diagram of numerical results displays this case when $\kappa_r$ is increased to 0.6. Continuing to increase $\kappa_r$, the frequency of M$_1$ will increase further and frequency of M$_2$ will still remain almost unaffected. As a result, this opens a bandgap associated with band inversion between $M_1$ and $M_2$ states. Due to the $C_{4v}$ symmetry of the bands, band inversion at M point will introduce ±1 Chern numbers exchange between the related bands. The band diagram of numerical results for $\kappa_r=0.8$ in Fig.1(c) confirms such a case in which band inversion takes place denoted by two bold bands with Chern numbers $C=\pm 1$ and topological bandgap shown by blue shading with $C_{gap}=+1$. Corresponding to this case, the TES (cyan line) appears in projected edge band diagram in Fig. 1(e) when the MPC in Fig. 1(c) is interfaced with a metallic boundary.[8,9] The insert in Fig. 1(e) plots the profile of eigenfield $E_z$ for the edge mode A which shows its "edge" property. To show the TES clearly, its one-way waveguide propagation at edge of the structure are simulated in Fig. 1(f). Here, the point source is denoted by a violet star with frequency at the horizontal line in Fig. 1(e). Metallic boundary is denoted by black bold line to prevent radiation loss into air.[8,9] The one-way waveguide propagations is consistent with the topological property of bandgap in Fig. 1(c), showing the validity of our perturbative approach.

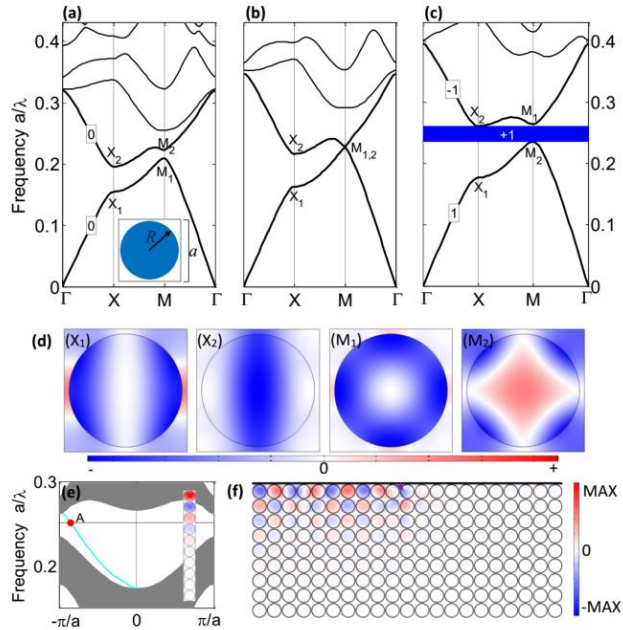

FIG. 1. (Color online) (a) Band diagram for a square lattice with $\kappa_r=0.4$, $\varepsilon=13$ and $R=0.46a$. Inset shows a unit cell. (b), (c) Band diagrams for lattice in (a) with $\kappa_r=0.6$ and $\kappa_r=0.8$, respectively. (d) Fields $2\mu_0 \text{Re}(iH_x^* H_y)$ for Bloch states $X_1$, $X_2$, $M_1$ and $M_2$ shown in (a). (e) TES (cyan line) in the projected edge band diagram. Inset: Profile of eigenfield $E_z$ for the edge state A. (f) One-way waveguide propagation field profile at edge of the MPC interfaced on the top with a metallic boundary.

The above designing employs the second term in Eq. (4) by altering the $\kappa_r$ of rods. Next, we give examples to engineer TESs using the first term in Eq. (4) by inserting isotropic rods (with relative permittivity $\varepsilon_i$). The microstructure we considered here is a 2D square lattice of magnetic rods with $R=0.13a$, $\kappa_r=0.4$ and $\varepsilon_r=13$. The band diagram of the structure is shown in Fig. 2(a). It is seen that there is no topological bandgap in such a structure within the frequency range from $a/\lambda=0.67$ to 0.8. Considering the topologies of the bands marked with each Chern number in Fig. 2(a), if we can separate the upper two bold bands enough to open a bandgap, the TES can be realized within this frequency range. Therefore as above example, we inspect the field-energy distributions $\varepsilon_0 |\mathbf{E}_{nk}(\mathbf{r})|^2$ in a unit cell at the related high symmetry Bloch points of $X_2$, $X_3$, $M_2$ and $M_3$ denoted in Fig. 2(a). The results are shown in Fig. 3.

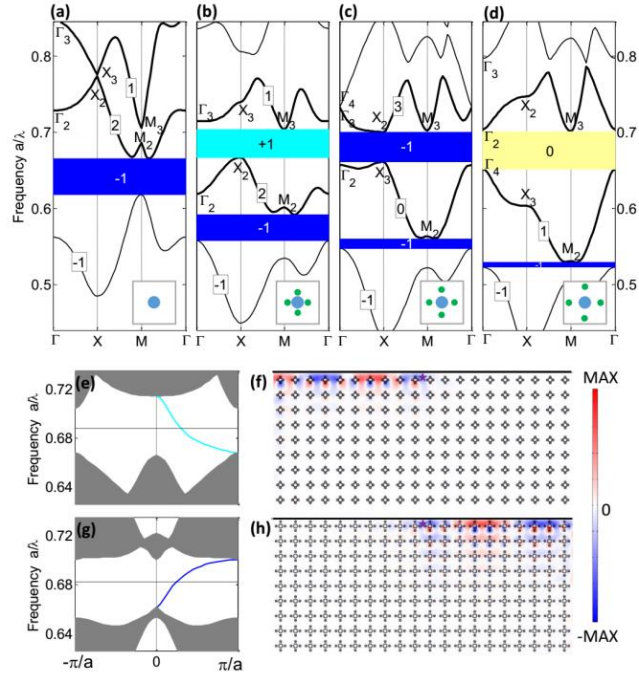

FIG. 2. (Color online) (a) Band diagram for a square lattice with $\kappa_r=0.4$, $\varepsilon=13$ and $R=0.13a$. (b)-(d) Band diagrams for lattices in (a) with symmetrically inserting isotropic rods ($\varepsilon_i=8.9$, $r_i=0.06a$ and $d=0.192a$, $0.292a$, $0.392a$, respectively) along $x=0$ and $y=0$ directions. In (a)-(d), insets show the unit cells. (e),(g) TES (cyan line and blue line) in the projected edge band diagram corresponding to (b) and (c), respectively. (f),(h) One-way waveguide propagation field profiles at edge of the MPC interfaced on the top with a metallic boundary.

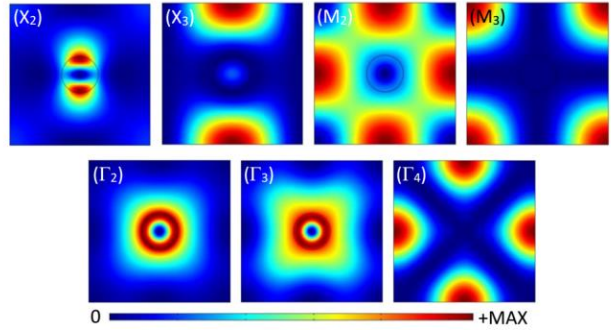

FIG. 3. (Color online) Fields $\varepsilon_0|\mathbf{E}_{nk}(\mathbf{r})|^2$ for the related Bloch states at $X_2$, $X_3$, $M_2$, $M_3$ and $\Gamma_2$-$\Gamma_4$ as shown in Fig. 2(a) (The $\Gamma_4$ point is not shown in Fig. 2(a)).

From Fig. 3, we observe that $\varepsilon_0|\mathbf{E}_{nk}(\mathbf{r})|^2$ is large for $X_2$ state near outside of the rod, but small for $X_3$ state. Thus, if we insert rods with $\varepsilon_i>1$ at these positions, the frequency of $X_2$ state will get a large decrease, while the frequency of $X_3$ will decrease small as can be seen from Eq. (4). For the same reason, if the insertions are near the boundaries of unit cell where $\varepsilon_0|\mathbf{E}_{nk}(\mathbf{r})|^2$ is large for state $X_3$ but small for state $X_2$, the opposite results can be obtained. Applying the same arguments to state $M_2$, we can deduce its frequency will decrease continuously as the insertions move out to the boundaries of unit cell along $x=0$ and $y=0$ directions. For state $M_3$, because of its near zero field along $x=0$ and $y=0$ directions, the insertions along these directions almost have no effect on its frequency.

According to the above analysis, we first symmetrically insert four isotropic rods near outside of magnetic rod along $x=0$ and $y=0$ directions (See insert in Fig. 2(b)) which will open the topological bandgap between the upper two bold bands. The band diagram of numerical results confirms such a case in Fig. 2(b) with inserted rods having radius $r_i=0.06a$ and $\varepsilon_i=8.9$ at positions with $d=0.192a$ away from the center. The opened topological bandgap is denoted by cyan shading with $C_{gap}=+1$, just equal to the sum of the Chern numbers of bands below it. Then we symmetrically change the insertions outward to positions with $d=0.292a$ away from the center (See inset in Fig. 2(c)). According to our perturbative prediction, in this case, the frequency of state $X_3$ will decrease more than that of $X_2$. As a result, the band inversion will take place between states $X_2$ and $X_3$. Considering the symmetry of the bands, this band inversion introduces ±2 Chern numbers exchange between the related bands, which will change the topology of opened bandgap. The band diagram of numerical results for this case shown in Fig. 2(c) confirms this prediction with opened bandgap denoted by blue shading with $C_{gap}$=-1 (the upper one), together with the Chern number of the band below the opened bandgap. At last, we move the insertions to positions near the boundaries of unit cell (See inset in Fig. 2(d)). Compared with the case in Fig. 2(c), in present, frequency of $M_2$ state will decrease and the separation between $X_2$ and $X_3$ will increase. The band diagram of numerical calculation with $d$=0.392$a$ in Fig. 2(d) agrees well with these estimations. But an important fact is that the frequency of $\Gamma_4$ state becomes lower than that of $\Gamma_2$ due to its large decrease. This behavior also can be understood with above analysis from the field-energy distributions $\varepsilon_0 |\mathbf{E}_{nk}(\mathbf{r})|^2$ at $\Gamma_2$, $\Gamma_3$ and $\Gamma_4$ states before insertions (See Fig. 3). The band inversion between $\Gamma_2$ and $\Gamma_4$ states introduces exchange of ±1 Chern numbers between them and leads to trivial topology of the opened bandgap as shown by light yellow shading in Fig. 2(d). For the constructed topological bandgaps in Figs. 2(b) and 2(c), the corresponding TESs, cyan line with negative slope and blue line with positive slope, are plotted in projected edge band diagrams in Figs. 2(e) and 2(g), respectively. And the one-way waveguide propagations supported by these TESs at edge of the microstructures are simulated in Figs. 2(f) and 2(h) with frequencies of point sources at the horizontal lines in Figs. 2(e) and 2(g), respectively. We see all these numerical simulations show good agreement with our perturbation analysis.

The above cases show the designs of single mode TESs. In the following, we give an example to engineer multimode TES by inserting isotropic rods. Figure 4(a) shows the band diagram of the considered microstructure composed of magnetic rods with $R$=0.1$a$, $\kappa_r$=0.45 and $\varepsilon_r$=13 in a square lattice. Within the frequency range from $a/\lambda$=0.7 to 1, there is no topological bandgap. If inserted rods can decrease the frequency of $\Gamma_3$ state more than $\Gamma_4$ and decrease $\Sigma_1$ more than $\Sigma_2$, bandgap between the upmost two bold bands can be opened. As before, in Fig. 4(c) we first display the field-energy distributions $\varepsilon_0 |\mathbf{E}_{nk}(\mathbf{r})|^2$ at these Bloch states. Based on the perturbative method of Eq. (4) and Fig. 4(c), symmetrically inserting four

isotropic rods near outside of the magnetic rod along x=±y directions (See insert in Fig. 4(b)) can open the topological bandgap. The band diagram of numerical results confirms this prediction in Fig. 4(b) with inserted rods having radius $r_i$=0.05$a$ and $\varepsilon_i$=8.9 at positions with $d$=0.152$a$ away from the center. Clearly within the frequency range from $a/\lambda$=0.82 to 0.88, a topological bandgap denoted by green shading with $C_{gap}$=+2 emerges in accordance with our perturbative estimation. The corresponding TESs (two green lines) are shown in Fig. 4(d) and the one-way waveguide propagation supported by them is simulated in Fig. 4(e) with the frequency at horizontal line in Fig. 4(d).

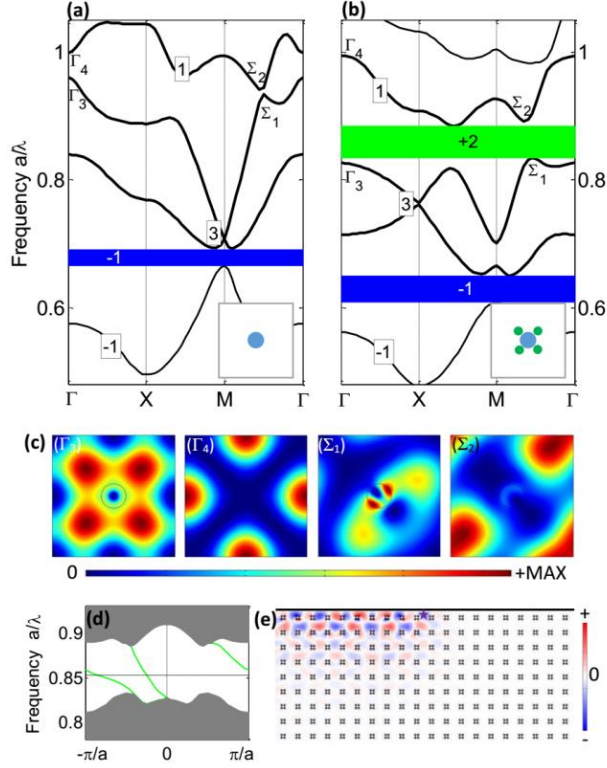

FIG. 4. (Color online) (a) Band diagram for a square lattice with $\kappa_r$=0.45, $\varepsilon$=13 and $R$=0.1$a$. (b) Band diagram for lattice in (a) with symmetrically inserting isotropic rods ($\varepsilon_i$=8.9, $r_i$=0.05$a$ and $d$=0.152$a$) along x=±y directions. In (a) and (b), insets shows the unit cells. (c) Field $\varepsilon_0 |\mathbf{E}_{nk}(\mathbf{r})|^2$ for Bloch states $\Gamma_3$, $\Gamma_4$, $\Sigma_1$ and $\Sigma_2$ shown in (a). (d) TES (green lines) in the projected edge band diagram. (e) One-way waveguide propagation field profile at edge of the MPC interfaced on the top with a metallic boundary.

In conclusion, we have proposed a simple and efficient method to engineer TESs in 2D MPCs. Through altering the parameters distributions of the microstructure, we demonstrate its validity in three concrete examples by exact numerical calculations. Our method is very general and it opens up a way to engineer TESs by altering the microstructures.

The authors gratefully acknowledge the financial support by the National Natural Science Foundation of China under Grant Nos. 11574031 and 61421001. And we would like to thank COMSOL Co., Ltd. for their COMSOL 5.1.